\documentclass[amsmath]{article}
\usepackage{graphicx}

\begin{document}

\author{K. Gatner \\ Faculty of Chemistry, University of Wroc\l{}aw, \\ 14
F. Joliot-Curie, 50-383 Wroc\l{}aw, Poland\\ A. N. Lachinov \\ Institute of Physics of Molecules and Crystals,\\ Russian Academy of Sciences, \\151 October Av., 450075 Ufa, Russia 
\\M. Matlak \footnote{Corresponding author: matlak@us.du.pl}, A. \'Slebarski \\ Institute of Physics, Silesian University,\\ 4 Uniwersytecka,
40-007 Katowice, Poland \\ and \\
T. G. Zagurenko \\ Institute of Physics of Molecules and Crystals,\\ Russian Academy of Sciences, \\151 October Av., 450075 Ufa, Russia }

\title{Fermi level and phase transformations in GdCo$_2$}
\maketitle
\begin{abstract}
We observed characteristic temperatures connected with different phase
transformations in the polycristalline, ferromagnetic $GdCo_2$ sample
possible to appear in the temperature range from $300$ $K$ to $440$ $K$. We
used here three independent methods applied to still the same $GdCo_2$
sample testing in this way the sensivity of each particular method. One of
them was the nonconventional, indirect measurement of the chemical potential
as function of temperature. In the second method we measured the current
flow through the $GdCo_2-polymer-reference$ $metal$ sandwich vs temperature
(surface potential barrier method). The first and second methods were
applied for the first time to the polycrystalline, magnetic $GdCo_2$
material. The third method was the measurement of the electric resistivity
vs temperature. These three independent measurements make an evidence that
within the measured temperature range plenty of transformations in the
sample take place and that the characteristic temperatures connected with
these transformations can be detected with the use of these methods
visualizing in this way very important role of the chemical potential in the
detection of phase transformations (transitions). The first method (indirect
measurement of the chemical potential vs temperature) revealed 24
characteristic temperatures $T_1-T_{24}$ connected with many different
transformation processes within the polycrystalline, ferromagnetic sample
under heating (reorientation, reordering processes, etc.), as well as, the
Curie temperature $T_C$ (second order phase transition:
ferromagnet-paramagnet). With the use of the second method (surface
potential barrier method) we could easily identify the characteristic
temperatures $T_8-T_{19}$ and $T_C$ in quite good agreement with the first
method. The second method revealed also several additional characteristic
temperatures which could be seen from appearing peaks in the measured
resistivity of the sandwich vs temperature but not seen when using the first
method. The application of the third method (the resistivity measurement vs
temperature) revealed almost the same values for the characteristic
temperatures $T_1-T_{14}$, $T_{18}-T_{24}$ and $T_C$ as found with the use
of the first method but several, additional characteristic temperatures
could also be detected which are not seen when apply the first method and
seen with the use of the second one. The existence of characteristic
temperatures, confirmed with the use of three independent measurements make
an evidence that plenty of transformations (transitions) really take place
when heating the polycristalline $GdCo_2$ sample. The applied methods
complement each other and can successfully be used to detect characteristic
temperatures of the system connected with different phase transformations.

PACS numbers: 64.60.-i, 75.30.Kz, 75.40.Cx \newline
Keywords: Magnetic alloys, chemical potential, phase transitions
(transformations).
\end{abstract}

%\maketitle

\section{\protect\smallskip Introduction}

It is commonly known that solids undergo phase transitions (first order,
second order) with changing temperature, characterized by a critical
temperature $T_C$ depending on the system (melting temperature, Curie (Neel)
temperature (magnetic, ferroelectric systems), superconducting transition
temperature, etc.). The temperature behaviour of real solids is, however,
very often much more complicated. There exist in reality many other
anomalies such as structural transformations, metastable or ''exotic''
phases, changes between different phases (metallic-nonmetallic and vice
versa, transitions between different magnetic or ferroelectric phases,
etc.). This very complicated picture of the transitions in real solids is
due to a great variety of different interaction processes in the coupled
ionic and electronic subsystems resulting in a variety of the observed
properties of solids. To each such a transition can usually be attributed a
characteristic temperature or characteristic temperature range where a
transition begins and ends.

This paper is devoted to a general problem how to detect experimentally all
critical and characteristic temperatures of solids taking as an example
polycrystalline, ferromagnetic $GdCo_2$ material and applying here three
independent experimental methods. A $GdCo_2$ ingot was prepared by a melting
of the constituent metals on a water-cooled copper hearth in a high purity
argon atmosphere. The sample was remelted several times to promote
homogeneity. The phase purity of the $GdCo_2$ compound was ascertained by
means of the X-ray Debye-Scherer diffraction with $Cu-K_\alpha $ radiation
using a Siemens D-5000 diffractometer. The intermetallic $GdCo_2$ sample,
prepared in this way, is polycristalline, metallic, ferromagnet with Curie
temperature $T_C$ of about $410$ $K$ (see e.g. Refs [1]-[6]) and this kind
of the prepared sample is especially suitable for our experiments. Except
the ferromagnetic phase transition at $T_C$ we can expect many other low
energetic transformation processes such as small displacements of $Gd$ and $%
Co$ ions within the unit cell, as well as, small relative movements of the
grain boundaries and domain structure with changing temperature what should
be reflected in our measurements indicating the existence of the
characteristic temperatures connected with these transformations. To
investigate the properties of this sample we apply a nonconventional method
of the indirect measurement of the chemical potential as function of
temperature (the galvanic cell method), surface potential barrier method
(measurement of the current flow vs temperature through the $GdCo_2$-\emph{%
polymer-metal} sandwich) and we perform the measurement of the temperature
dependence of the resistivity (cf also Ref. [6]). It should be stressed at
this place that the galvanic cell method and the surface potential barrier
method were applied for the first time to this material in the present
paper. All three independent measurements, applied to the same
polycristalline $GdCo_2$ sample, reveal plenty of characteristic
temperatures, connected with mentioned transformation processes, as well as,
the Curie temperature connected with the magnetic phase transition of the
system. The use of three independent methods was absolutely necessary to
confirm the existence of the observed characteristic temperatures as a
physical reality and not as artifacts or measurement errors. Taking into
account that each particular, experimental method possesses its own
characteristic sensitivity the agreement between characteristic temperatures
obtained from three independent measurements is quite good. It, however,
means that the characteristic temperatures connected with different phase
transformations (transitions) can successfully be detected with the use of
these methods with sufficient accuracy.

\section{Galvanic cell method}

One of the methods applied in this paper is based on the fact that the
chemical potential of the system can ''see'' both critical and
characteristic temperatures of solids. First observations of this kind were
made for superconductors, Refs [7]-[16], both theoretically and
experimentally. Further theoretical investigations, Refs [17]-[24], revealed
that the temperature dependence of the chemical potential shows kinks for
critical and characteristic temperatures also in the case of magnetic and
structural transitions for both normal and fluctuating valence systems. In
the case of alloys, when a transition appears for a critical concentration,
the chemical potential can also ''see'' this transition, too (cf Refs [18],
[21]). As a generalization of these results it was suggested that the
chemical potential should experimentally be used to detect all transitions
and transformations possible to occur in real solids with changing
temperature or concentration (see Refs [18]-24]). The realization of this
idea is, however, not an easy problem and therefore only a few experimental
measurements have been performed but all of them support this general idea.
In Refs [12]-[14], [16] the chemical potential as function of temperature
has been measured for superconductors with the use of the work function
method. Additionally, in Ref. [14] has been found that the chemical
potential can also ''see'' the magnetic transitions in thin films when apply
external magnetic field. Recently, in Refs [25], [26], a quite new approach
for the indirect measurement of the chemical potential of metallic samples
has been proposed with the use of a simple electrochemical experiment,
utilising a galvanic cell. One of the electrodes of this cell was the
investigated metallic sample and the other one the reference electrode.
According to the Nernst's formula (cf e.g. Ref. [27]) the difference of the
chemical potentials of the electrodes is given by the expression

\begin{equation}
\mu ^{(sample)}-\mu ^{(ref)}=c(\Phi ^{(ref)}-\Phi ^{(sample)})
\end{equation}
where $c$ is a constant and $\Phi ^{(ref)}-\Phi ^{(sample)}$ is the measured
voltage. By changing the temperature of the electrolyte bath the voltage as
function of temperature can be plotted, giving us indirectly the relative
chemical potential temperature dependence. Using as an electrode
investigated magnetic samples $Gd$, $Cr$ (see Ref. [25]) and $Gd_5Si_4$ (see
Ref. [26]) it was possible to identify the critical temperatures of these
materials directly from the temperature dependence of the voltage in
agreement with existing experimental data. Thus, the method can visualize
magnetic phase transitions taking place in the investigated cell electrode
by changing the temperature of the electrolyte bath. In the case of $%
Gd_5Si_4 $ material (cf Ref. [26]) it was also possible to detect a
characteristic temperature $T^{\star }$ connected with the inhomogeneity of
the measured sample. This method applied to the shape memory alloys $TiNi$
(10\% of deformation) in Ref. [25] and $TiNi$ (15\% of deformation) in Ref.
[26] revealed plenty of characteristic temperatures connected with
structural transformations (begin and the end of each transformation) in
agreement with the known experimental results. The same can be said about $%
CuAlNiTiMn$-alloy sample (see Ref. [26]) where the temperature of the
structural transformation was easy to identify and its value was equal to
the value known independently from another experiment. In other words the
method works and indicates all possible critical and characteristic
temperatures of the measured system. The only problem is how to overcome a
relatively narrow temperature range available for the experiment
(electrolyte) and how to adopt the method to non-metallic samples. The
situation can, however, rapidly be changed in the nearest future due to new
investigations.

In the case of the $GdCo_2$ sample with relatively high Curie temperature a
special electrochemical cell with ionic liquid was designed:

\begin{equation}
\begin{array}{llll}
(-)GdCo_2\mid & S1\mid\mid & S2\mid & Ag(+),
\end{array}
\end{equation}
\[
S1=CoCl_2(2\%) +EACl(20\%)+DEACl(78\%),
\]
\[
S2=AgCl +EACl +DEACl
\]

%\begin{equation}
%\begin{array}{llll}
%(-)GdCo_2\mid\mid&
%\begin{array}{l}
%CoCl_2(2\%) \\
%+EACl(20\%) \\
%+DEACl(78\%)
%\end{array}
%&  \mid
%\begin{array}{l}
%AgCl \\
%+EACl \\
%+DEACl
%\end{array}
%&\mid\mid Ag(+)
%\end{array}
%\end{equation}

where $EACl$ is ethylammonium chloride and $DEACl$ is diethylammonium
chloride. The mixture of both organic salts and cobalt chloride yields
electrolyte, liquid above 71 C (344 K) and therefore it is possible to
perform the experiment to temperatures higher than the Curie temperature of
the sample ($T_C=410$ $K$). The reference electrode was silver-silver
chloride (2\%) inside of the same organic melting alkalammonium chloride,
separated with a glass frit from the electrolyte. It was established that
the resulting $EMF$ changes of the galvanic cell (2) do not depend on the
use of the reference electrode. The $EMF$ depends, however, on the
reactivity of the $Co$ ion in the electrode material and therefore it is
sensitive to the chemical potential changes due to the phase transition at $%
T_C$ and another transformations. The temperature sensor, used in the
experiment, was Pt-100. The $EMF$ of the cell and the temperature were
recorded with the use of the multichannel electrochemical set EMU (Elchema,
Wroclaw, Poland) connected with PC. The investigated $GdCo_2$ sample was
soldered with indium to copper wire and insulated with epoxy resin in such a
way that only the measured sample had contact with the electrolyte. Heating
and cooling curves of the $EMF$ as function of temperature were recorded.
The temperature scan was about 2 degrees/minute.

The results of the measurements can be seen in Fig. 1 (heating) and in Fig.
2 (cooling). The curve in Fig. 1, showing the temperature dependence of the $%
EMF$ of the galvanic cell, is not smooth. It visualizes many characteristing
temperatures (denoted by $T_1$, $T_2$, .. , $T_{24}$; $T_m$ is the melting
temperature of the electrolyte and $T_C$ is the Curie temperature). All of
them correspond either to a local minimum (maximum) or to a change in slope
of the $EMF$ curve. Such characteristic points in the $EMF$ curve always
signalize that the measured system undergoes a transformation (transition)
at these points (cf Refs [25], [26]). At the first glance, the
characteristic temperatures may look like artifacts or measurement errors.
It is, however, absolutely not the case because the appearance of these
points strongly confirm our further measurements. These points make an
evidence that with increasing temperature many different processes inside of
the sample take place (as mentioned above), including also the magnetic
phase transition at $T_C$. In other words, a real material transforms itself
with increasing temperature and certainly such a feature holds for another
materials, too, more or less. Surprisingly, in Fig. 1 we can easily find a
point corresponding to the melting temperature of the electrolyte $T_m=344$ $%
K$ and below this temperature we can find such characteristic temperatures
as $T_1-T_8$, confirmed also in our resistivity experiment (see later). In
other words the galvanic cell seems to work in a temperature range which is
lying below the melting point of the electrolyte. In contrary to Fig. 1, the
curve of the $EMF$ vs temperature presented in Fig. 2 (cooling), is much
more smooth and only several characteric temperatures $T_{10}$, $T_{20}$, $%
T_{22}$, $T_{23}$ and $T_{24}$ (a little shifted) can easily be identified
together with the Curie temperature $T_C$ , as well as, the melting
temperature of the electrolyte $T_m$ when compare their values with the
values presented in Fig. 1. This was certainly caused by irreversible
reorientation processes in the $GdCo_2$ sample after performed heating up to
temperatures higher than the Curie temperature. A more detailed discussion
of all characteristic temperatures in connection with the results of two
further measurements will be presented in Sect. 5.

\begin{figure}[h]
\begin{center}
\noindent
\includegraphics[angle=-90,scale=0.30]{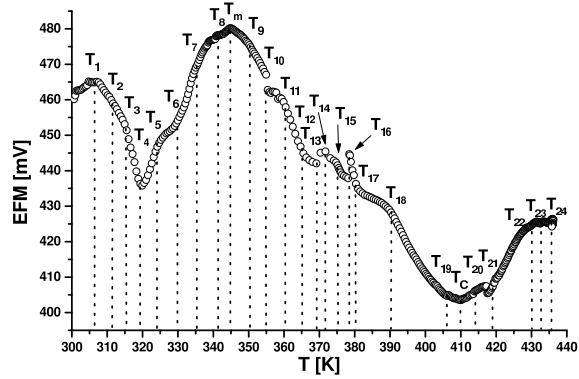}
\caption{{{{{\protect\footnotesize {Plot of the measured $EMF$ (multiplied by (-1)) vs temperature for $%
GdCo_2$ sample under heating. $T_M$ is the melting temperature of the
electrolyte, $T_1-T_{24}$ are the characteristic temperatures, $T_C$ is the
Curie temperature.}}}}}}
\end{center}
\end{figure}

\begin{figure}[h]
\begin{center}
\noindent
\includegraphics[angle=-90,scale=0.25]{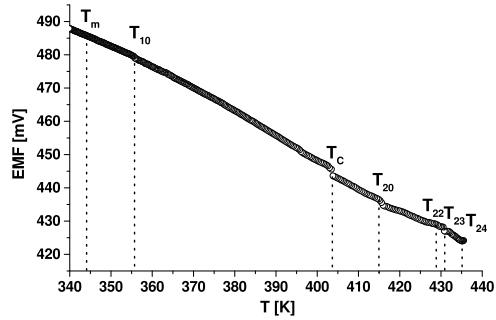}
\caption{{{{{\protect\footnotesize {The same as in Fig. 1 but under cooling with characteristic
temperatures $T_{10}$, $T_{20}$, $T_{22}$, $T_{23}$, and $T_{24}$. $T_m$ is
the melting temperature of the electrolyte and $T_C$ is the Curie
temperature.}}}}}}
\end{center}
\end{figure}

%fig. 1. Plot of the measured $EMF$ (multiplied by (-1)) vs temperature for $%
%CdCo_2$ sample under heating. $T_M$ is the melting temperature of the
%electrolyte, $T_1-T_{24}$ are the characteristic temperatures, $T_C$ is the
%curie temperature.

%fig. 2. The same as in Fig. 1 but under cooling with characteristic
%temperatures $T_{10}$, $T_{20}$, $T_{22}$, $T_{23}$, and $T_{24}$. $T_m$ is
%the melting temperature of the electrolyte and $T_C$ is the Curie
%temperature.

\section{Surface potential barrier method}

Several years ago a new experimental technique to study phase transitions in
metals has been found, called surface potential barrier method. This
technique has successfully been applied to study melting-crystallization
processes of low temperature melting metals like $Sn$, $In$ and Wood alloy
(cf Ref. [28]). Later on the method was utilized to investigate structural
transformations in nanostructured metals subjected by a severe plastic
deformations in $Cu$ (see Ref. [29]), $Ni$ (see Ref. [30]) and in the shape
memory alloys $TiNi$ (cf Ref. [31]). The method works in the following way.
A tunnel transparent potential barrier between an investigated metal ($IM$)
and a probe material ($PM$) can easily be formed due to an extreely narrow
conduction band (conduction level) lying between the valence and conduction
band of the $PM$ material. Thus, under applied voltage , the maximum of the
current flow through the barrier can be reached in the case when the Fermi
levels of $IM$ and $PM$ coincide. In the region of temperatures where $IM$
exhibits a phase transition that energy balance is broken due to a small
shift of the $IM$ Fermi level. This, however, leads to an exponential change
in the tunnel current flowing through the barrier which can be detected (cf
Ref. [32]) as an evidence for a phase transition (transformation). In other
words, the method uses also the fact that the chemical potential (Fermi
level in metals) ''feels'' the phase transition (transformation), similarly
to the galvanic cell method, described above. As a $PM$ material an
electroactive polymer was used.

To realize the surface potential barrier an experimental cell has been made
in the form of a multilayer sandwich type metal-polymer-metal structure. One
of the metals in the cell was the investigated metal ($IM$), exhibiting a
phase transition (transformation) lying within the studied temperature
interval. The other one, such a reference metal ($RM$) has been chosen which
does not exhibit a phase transition (transformation) in the investigated
temperature range. A polymer film, lying between $IM$ and $RM$, was made by
a spin coating technique of the polymer solution applied to the polished $IM$
surface. The $RM$ material was subsequently evaporated onto the free polymer
surface.

As a $PM$ polymer the poly(phthali\-dilideneby\-phenililene) material ($PPB$)
was used, similarly to the Ref. [33]. The following reasons justify this
choice. First, the polymer possesses very good film forming properties on
metallic substrates and as it was shown in Ref. [34] in certain conditions $%
PPB$ forms solid homogeneous films (0.005-10 $\mu m$ thick). Second, the $%
PPB $ material does not change the properties till the softening temperature
(about $710$ $K$ in vacuum). Third, the high conductive properties of this
polymer has been confirmed by an extensive study, performed in Ref. [35].
The polymer films of 0.9 $\mu m$ thickness were used in the experiment. The
quality of the polymer films was controlled by the scanning probe microscopy
method, Ref. [36]. The sandwich with $IM-PM-RM$ structure was placed into a
thermostat which was able to change the temperature within prescribed range
of 300-475 $K$. The temperature was varied with a velocity from 4 to 27
degrees/minute. The precision of the temperature measurement was 0.5 $K$.

The sample was connected to the electrical circuit which is typically used
for non-linear elements with $S$-shape negative differential resistance (see
Ref. [37]). A current source was used as a power supply. The current flowing
through the sandwich has not exceeded the value of 50 $mA$. The precision of
the current measurement was 5\%. The voltage drop on the sandwich was
normally from 1 $mV$ to 5 $V$.

\begin{figure}[h]
\begin{center}
\noindent
\includegraphics[angle=-90,scale=0.25]{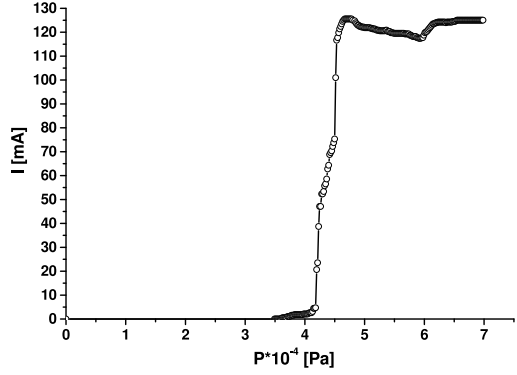}
\caption{{{{{\protect\footnotesize {The current-pressure dependence of the
metal-polymer-metal system. }}}}}}
\end{center}
\end{figure}

%Fig. 3. The current-pressure dependence of the metal-polymer-metal system.

\begin{figure}[h]
\begin{center}
\noindent
\includegraphics[angle=-90,scale=0.25]{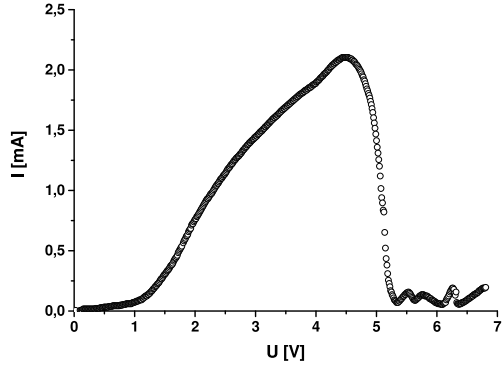}
\caption{{{{{\protect\footnotesize {The current-voltage characteristics of
the metal-polymer-metal system.}}}}}}
\end{center}
\end{figure}

%Fig. 4. The current-voltage characteristics of the metal-polymer-metal
%system.

To produce extremally narrow conduction band lying within the gap between
the valence and conduction band of the $PPB$ polymer the following procedure
has been applied. The high conducting state of the $PM$ layer was induced by
the unuaxial pressure which was applied to the multilayer sandwich type $%
IM-PM-RM$ structure. In Fig. 3 we show the conductivity dependence of the $%
PM $ vs applied pressure. It can be seen that the initial conductivity was
small and when the pressure reached a certain value, $P_{cr}$, the
conductivity started to increase and reached a relatively high value typical
for the metallic state and possessed metal-like temperature dependence down
to 1 $K$ (cf Ref. [38]).

\begin{figure}[h]
\begin{center}
\noindent
\includegraphics[angle=-90,scale=0.25]{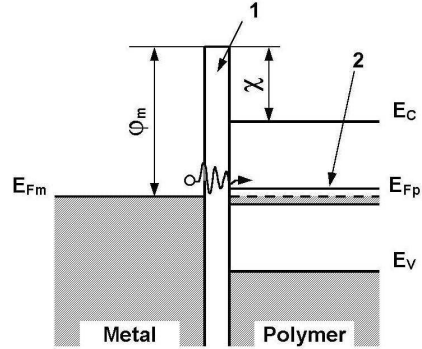}
\caption{{{{{\protect\footnotesize {Energy diagram of the metal-polymer-metal interface, the case of
flat bands. 1 - tunnel barrier; 2 - induced conduction level; $E_{Fm}$ and $%
E_{Fp}$ - the Fermi level of the metal and the polymer, respectively; $%
\varphi _m$ - the work function of the metal; $\chi $ - the electron
affinity of the polymer; $E_C$ and $E_V$ - energy of the conduction band
bottom and the valence band top of the polymer, respectively.}}}}}}
\end{center}
\end{figure}

%Fig. 5. Energy diagram of the metal-polymer-metal interface, the case of
%flat bands. 1 - tunnel barrier; 2 - induced conduction level; $E_{Fm}$ and $%
%E_{Fp}$ - the Fermi level of the metal and the polymer, respectively; $%
%\varphi _m$ - the work function of the metal; $\chi $ - the electron
%affinity of the polymer; $E_C$ and $E_V$ - energy of the conduction band
%bottom and the valence band top of the polymer, respectively.

The current-voltage ($I-V$) characteristics of such a multilayer structure
(sandwich) is presented in Fig. 4. This characteristics is typical for
systems with tunnel potential barrier. An analogical procedure was also used
to produce a tunnel barrier in the metal-$PPB$ system to obtain a Josephson
junction, Ref. [39]. The energetic diagram for the $IM-PM$ interface is
presented as a result of the narrow conduction band forming in the polymer
with a tunnel potential barrier within the contact area (see Fig. 5).

\begin{figure}[h]
\begin{center}
\noindent
\includegraphics[angle=-90,scale=0.30]{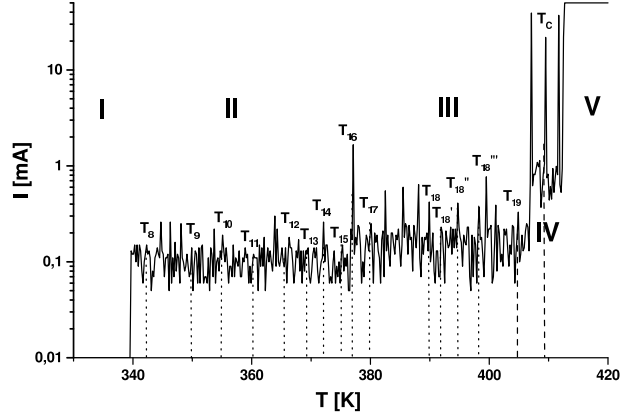}
\caption{{{{{\protect\footnotesize {\ Current (in the logarithmic scale) vs
temperature of the $GdCo_2$ $-PM-RM$ system for applied voltage $5V$.}}}}}}
\end{center}
\end{figure}

%Fig. 6. Current (in the logarithmic scale) vs temperature of the $%
%GdCo_2-PM-RM$ sandwich for applied voltage 5 $V$.

In Fig. 6 we show the temperature dependence of the current flowing through
the $GdCo_2-PM-RM$ sandwich. Five specific temperature regions can be
distinguished in this dependence. The first one corresponds to the low
conducting state of the system and observed up to $T$=339 $K$. In this
temperature range the signal is to small to detect phase transformations.
The current starts to increase at above 339 $K$ and this is the beginning of
the second region where a growth of the current fluctuations is observed. In
average, both the current values and current fluctuation amplitudes change a
little up to about 375 $K$.

\begin{figure}[h]
\begin{center}
\noindent
\includegraphics[angle=-90,scale=0.25]{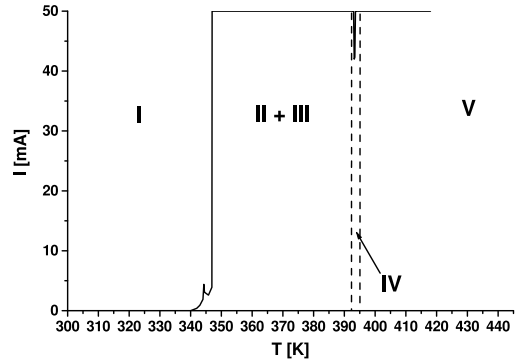}
\caption{{{{{\protect\footnotesize {The same in Fig. 6 but for applied
voltage 0.1 $V.$}}}}}}
\end{center}
\end{figure}

%Fig. 7. The same as in Fig. 6 but for applied voltage 0.1 $V$.

A next step in the current flow was observed beneath 380 $K$. This step
corresponds to the third teperature region. The average current amplitudes
and fluctuations in the third region are approximately two times greater
than in the second region.

Third and fifth region in Fig. 6 are separated by a narrow temperature range
(5 $K$) of abnormally big fluctuations (fourth region) with increasing
amplitudes and frequency fluctations till the boundary of the fifth region
is reached at approximately 410 $K$ (Curie temperature). The current reaches
its maximum value in the fifth region and the fluctuations disappear. Under
cooling the current fluctuations in the sandwich appear below 400 $K$ and
current drops to a very small value. As a rule, the temperature of the drop
is less than the Curie temperature of about 2 - 5 $K$. This histeresis may
be connected with irreversible reorientation processes of the
polycristalline sample accompanied by the change in its magnetic structure
(domain structure) after previously performed heating. Another types of
peculiarities in the current behaviour with decreasing temperature were not
observed.

We see that the value of the critical temperature $T_C$ of the magnetic
phase transition in the $GdCo_2$ sample, observed with the use of the
surface potential barrier method, agrees pretty well with the existing
experimental data (cf Refs [1]-[6]) and the value obtained from Fig. 1. It
is also important to stress that the characteristic temperatures $T_8-T_{19}$
are visible in Fig. 6, similarly to presented in Fig. 1, while the others ($%
T_1-T_7$) are hidden in the very small values of the current signal in the
first region (see Fig. 6). Within the third region in Fig. 6 the
characteristic temperatures $T_{18}^{\prime }$, $T_{18}^{\prime \prime }$
and $T_{18}^{\prime \prime \prime }$ (not resolved in Fig. 1) are quite good
visible in agreement with the resistvity measurements, presented in Fig. 8.

Within the framework of the potential barrier method the following studies
has been performed with respect to the repeatibility of the main data, the
influence of the temperature change rate on $I(T)$ dependiences, as well as,
to the influence of the applied voltage on $I(T)$. To this purpose more than
50 control measurements of the heating - cooling cycles were performed. The
estimation of the credibility of the measured data is 95 \%. Thus, it is
more than enough to exclude incidentality.

The velocity of the temperature change was 3 - 27 $K$/min in the whole
measured temperature interval. It was established that there was no
influence to the amplitudes of the registered features and to their
positions on the temperature scale. The most changes in the $I(T)$
dependence took place only when decrease the voltage applied to the
experimental cell. The voltage was changed in the interval from 1 $mV$ - 5 $%
V $. The choice of the highest voltage limit was restricted by the shape of
the $I-V$ characteristics (see Fig. 4) because the measurements should take
place only on the linear part of the curve. The application of the lowest
voltage limit was connected with the equipment parameters.

\begin{figure}[h]
\begin{center}
\noindent
\includegraphics[angle=-90,scale=0.30]{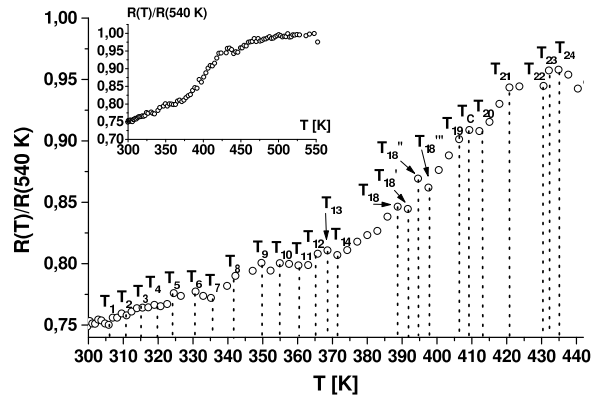}
\caption{{{{{\protect\footnotesize {Resistivity vs temperature for $GdCo_{2}$ sample.
The insert shows the resistivity in the wider temperature range.}}}}}}
\end{center}
\end{figure}

%Fig. 8. Resistivity vs temperature for $GdCo_2$ sample. The inset shows the
%resistivity in a wider temperature range.

Fig. 7 shows the temperature dependence of the current through the $%
GdCo_2-PM-RM$ sandwich at the applied voltage of 0.1 $V$. The comparison of
the curves presented in Fig. 6 and Fig. 7 shows the same features in the
temperature range near the magnetic transition. The main changes are only
visible in the temperature areas between both the first and second, as well
as, between the second and the third regions. We see that the begin of the
second region is shifted to higher temperatures of about 7 $K$. The typical
current step between the second and the third area disappears (becomes
invisible on the fluctuation amplitude level). It is quite naturally that
the decrease of the voltage leads to the decrease of the measuring current,
as well as, to the decrease of the fluctuation amplitudes. This is a trivial
result of the Ohm's law. That is why the fine structure of phase
transformations near the boundary of the second and third areas are weakly
resolved. On the other hand the applied voltage can influence the surface
potential barrier parameters such as shape and width of the barrier. For
example, the width of the barrier should be larger at the lower voltage
value than at the higher one. It is also well known that the tunneling
probability decreases when the potential barrier width increases. The shift
of the first low temperature step to the highest temperature region can be
explained as follows. The increase of the barrier width leads to the
decrease of the measured sensitivity to the small energetic perturbations
onto the metal-polymer ($IM-PM$) interface. Therefore the pretransition
fluctuations were recorded later on the temperature scale than in the case
when the applied voltage was greater and the beginning of the curve took
place in the area of highest pretransition energetic perturbations. However,
it can be seen that the positions of the lower temperature edges almost
coincide in both curves (Fig. 6 and Fig. 7).

\section{Resistivity measurement}

The electrical measurement applied to the same $GdCo_2$ sample was performed
in the temperature range between 300 $K$ and 500 $K$ using a standard four
lead ac method. The result of this measurement can be seen in Fig. 8. The
curve possesses structure with many characteristic temperatures indicating
plenty of transformations when heating the sample. All characteristic
temperatures $T_1-T_{14}$, $T_{18}-T_{24}$ and the Curie temperature $T_C$
in this curve coincide quite good with the previous measurement, depicted in
Fig. 1. The temperatures $T_8-T_{14}$, $T_{18}$, $T_{19}$ and $T_C$ agree
pretty well also with the result of the potential barrier method, presented
in Fig. 6. The charecteristic temperatures $T_{15}-T_{17}$, resolved in Fig.
1, are not seen in Fig. 8. We can, however, see that between $T_{14}$ and $%
T_{18}$ in Fig. 8 a very small instability occurs lying on the resolution
limit of the resistivity measurement what signalizes only possible
transitions. Between characteristic temperatures $T_{18}$ and $T_{19}$ in
Fig. 8 we can find the temperatures $T_{18}^{\prime }$, $T_{18}^{\prime
\prime }$ and $T_{18}^{\prime \prime \prime }$, seen also in Fig. 6 but not
detected in Fig. 1.

When try to theoretically decribe the shape of the resistivity curve we
should apply a suitable microscopic model for $GdCo_2$ and use e.g. Kubo
formula for the electric conductivity [40]. This formula, however, contains
a current-current thermal average over the Gibbs great canonical ensemble
with the chemical potential inside (current-current correlation function).
Thus, because the chemical potential is sensitive to phase transformationsand magnetic transition at $T_C$ (see Fig. 1) the restivity, calculated with
the use of a suitable microscopic model and Kubo formula should also reflect
all characteristic temperatures of the system via the chemical potential
temperature dependence. In other words, it should be clear that, in fact,
the resistivity curve should also be sensitive to different phase
transformations and transitions and it is really the case (see Fig. 8).

\section{Conclusions}

We have used three independent methods to detect characteristic temperatures
connected with different phase transformations (transitions) possible to
appear when heating $GdCo_2$ material sample. It is important to stress that
in all three measurements exactly the same sample was used. Because of the
polycristalline and magnetic nature of the investigated material
characteristic temperatures should reflect all the mentioned processes
inside the sample under heating, as well as, the magnetic transition at $T_C$%
. The most important results of the paper, presented in Fig. 1, Fig. 6 and
Fig. 8, entirely support this general statement. We have, however, to take
into account that each particular method, applied in this paper, possesses
its own limitations (sensititvity, resolution depending on temperature,
etc.) and therefore we cannot expect the agreement between the methods to be
100\%.

The indirect measurement of the chemical potential revealed the existence of
the characteristic temperatures $T_1-T_{24}$ and $T_C$ connected with the
characteristic points in Fig. 1. These points corresponded to kinks, change
in slope and local maxima (minima) of the $EMF$ curve. It is important to
stress that the characteristic temperatures $T_1-T_8$ where detected below
the melting point of the electrolyte $T_m$ (see Fig. 1) and their values
coincided quite well with the instabilities of the resistivity curve,
depicted in Fig. 8. It makes an evidence that the $EMF$ measurement can also
be useful in the temperature range lying below the melting temperature of
the electrolyte, used in the experiment. The characteristic temperatures $%
T_1-T_{14}$, $T_{18}-T_{24}$ and the Curie temperature $T_C$ (of about 410 $%
K $ (see e.g. Refs [1]-[6])), found in Fig. 1, correspond quite well with
the irregularities in the resistivity curve in Fig. 8. There were, however,
characteristic points $T_{15}-T_{17}$, resolved in Fig. 1, but hardly seen
in Fig. 8 because the irregularities in the temperature range from $T_{14}$
to $T_{18}$ were so small that they could be interpreted (when looking only
in Fig. 8) as a measurement error. The characteristic points $T_{15}-T_{17}$
were, however, resolved in Fig. 6. In the temperature range between $T_{18}$
and $T_{19}$ in Fig. 1 the $EMF$ curve is smooth. It does not indicate at
all the existence of the characteristic temperatures $T_{18}^{\prime }$, $%
T_{18}^{\prime \prime }$ and $T_{18}^{\prime \prime \prime }$, very well
seen in Fig. 6 and Fig. 8. May be that the $EMF$ measurement was not
sensitive enough to resolve these characteristic temperatures. The same can
be said about the temperatures corresponding to many peaks in Fig. 6, not
resolved in Fig. 1 and Fig. 8. We should, however, to be aware of the fact
that from three methods applied here a more 'static' was the $EMF$
measurement. The current flow through the $GdCo_2$ sample, recorded when
using the potential barrier method or resistivity measurement can, in
principle, ''induce'' (as a ''dynamic'' effect) small structural changes in
the sample during the mesurement process (influence grain boundaries, domain
structure, etc.) which manifested itself in a form of additional peaks in
Fig. 6 (not resolved in Fig. 1 and Fig. 8) and additional characteristic
points ($T_{18}^{\prime }$, $T_{18}^{\prime \prime }$ and $T_{18}^{\prime
\prime \prime }$) in Fig. 6 and in Fig. 8, not seen in the $EMF$ measurement
(Fig. 1). A more general conclusion resulting from three independent
measurements is certainly this that the characteristic points $T_1-T_{24}$
and $T_C$ from Fig. 1 cannot be interpreted as artifacts because they appear
as a result of three independent measurements. Unfortunately, due to very
small signal when using the potential barrier method in the temperature
range below 340 $K$ the charecteristic temperatures $T_1-T_7$, present in
Fig. 1, are confirmed only by the resistvity measurement in Fig 8. The
methods complement each other and indicate transformation (transition)
temperatures of the measured system. They say, however, nothig about the
origin of a given transformation (transition). To obtain a more detailed
information what really happens at a given characteristic temperature we
should apply another experimental methods (it lies outside of the scope of
the present paper). The information about transition temperatures of the
system is, however, very useful indication because we exactly know in which
temperature range another additional experiments should be performed to find
the origin of the transformation (transition).

Three methods applied in this paper possess something in common. All of them
visualize very important role of the chemical potential in phase
transformations (transitions) of real solids as suggested in Refs [17]-[26].
We have shown that at characteristic temperatures the chemical potential of
the electronic subsystem of the investigated material changes in such a way
that we can record it as a physical reality and interpret it in three
independent measurements (Fig. 1, Fig. 6, Fig. 8) as a signal of the
arriving phase transformation (transition). Taking as an example magnetic $%
GdCo_2$ sample we have also demonstrated that the heating process of solids
(especially policristalline) is by no means a simple and smooth process
because except for the magnetic transition at $T_C$ the system transits many
other phase transformations, mentioned above.

\emph{Acknowledgements}

We are grateful to Prof. W. Suski for critical remarks concerning our
results.

\end{document}